\documentstyle[12pt]{article}
\newcommand{\be}{\begin{equation}}
\newcommand{\bea}{\begin{eqnarray}}
\newcommand{\eea}{\end{eqnarray}}
\newcommand{\ba}{\begin{array}}
\newcommand{\ea}{\end{array}}
\newcommand{\ee}{\end{equation}}

\expandafter\ifx\csname mathbbm\endcsname\relax

\else

\fi
\begin{document}
\begin{titlepage}
\hfill
\vbox{
    \halign{#\hfil         \cr
           IPM/P-2002/024 \cr
           hep-th/0206237  \cr
           } 
      }  
\vspace*{20mm}
\begin{center}
{\Large {\bf Orbiting Membranes in M-theory on $AdS_7\times S^4$
Background}\\ }

\vspace*{15mm}
\vspace*{1mm}
{Mohsen Alishahiha$^{a}$  and  Masumeh Ghasemkhani$^{b}$}\\

\vspace*{1cm}

{\it$^a$ Institute for Studies in Theoretical Physics and Mathematics (IPM)\\
P.O. Box 19395-5531, Tehran, Iran\\
$^b$Department of Physics, Alzahra University, Tehran 19894,
Iran \\}

\vspace*{1cm}
\end{center}

\begin{abstract}
We study classical solutions describing rotating and boosted membranes
on $AdS_7\times S^4$ background in M-theory. We find the dependence of
the energy on the spin and R-charge of these solutions. 
In the flat space limit we get $E\sim S^{2/3}$, while for
$AdS$ at leading order $E-S$ grows as $S^{1/3}$. The membranes
on $AdS_4\times S^7$ background have briefly been studied as well. 

\end{abstract}  

\end{titlepage}

\newpage

\section{Introduction}
By now it is believed that type IIB
string theory on $AdS_5\times S^5$ with $N$ fluxes on the 
$S^5$ is dual to four dimensional ${\cal N}=4$ $SU(N)$ SYM 
theory \cite{{MAL},{GKP1},{WI}}. According to this conjecture 
the spectrum of single string state on $AdS_5\times S^5$ 
corresponds to spectrum of single trace operators of the 
${\cal N}=4$ gauge theory. However, until recently, this 
correspondence had only been studied for supergravity modes 
on $AdS_5\times S^5$ which are in one-to-one correspondence 
with the chiral operators of the ${\cal N}=4$ gauge theory. 
Basically this is because formulating the appropriate
sigma models and solving them for such a curved background 
with RR-field is not an easy work. On the other hand the 
massive string modes correspond to operators in long multiplets 
whose dimensions grow  as $\lambda^{1/4}$ for large `t Hooft 
coupling $\lambda$. Fortunately one can get rid of  
these problems by considering those operators with very high 
bare dimension, such as operators with high R-charge, spin, etc. 
Practically it has been shown in \cite{{BMN}, {GKP}} that this 
procedure works for a special class of operators 
with high R-charge or high spin. 

In fact it has been conjectured \cite{BMN} that string theory on 
the maximally supersymmetric ten-dimensional PP-wave has a 
description in terms of a certain subsector of the
large $N$ four-dimensional ${\cal N}=4\; SU(N)$ supersymmetric 
gauge theory at weak coupling. More precisely this subsector is 
parametrized by states with conformal weight $\Delta$ carrying 
$J$ units of charge under the $U(1)$ subgroup of the $SU(4)_{R}$ 
R-symmetry of the gauge theory, such that both $\Delta$ and $J$ 
are parametrically large in the large `t Hooft coupling while 
their difference, $\Delta-J$ is finite. Then it has been possible 
to work out the perturbative string spectrum from gauge theory 
side. The idea of \cite{BMN} is based on the observation that 
the PP-wave background with RR 4-form in type IIB string theory 
is maximally supersymmetric \cite{Blau1}, solvable \cite{Met, MT} 
and can be understood as a certain limit (Penrose limit) of 
$AdS_5\times S^5$ geometry \cite{{Blau2},{BMN}}. Although in the 
BMN consideration the background is maximally supersymmetric,
it has been shown that this conjecture can also be applied for
the backgrounds with less supersymmetry \cite{all}.

In an attempt to explore this idea the authors of \cite{GKP} 
identified certain classical solutions representing highly 
excited string states carrying large angular momentum in the 
$AdS_5$ part of the metric with gauge theory operators with high 
spin $S$ and conformal dimension $\Delta$ which is identified  
with the classical energy of the solution in the global
$AdS$ coordinates. An interesting observation of \cite{GKP} is 
that the classical
energy of the rotating string in $AdS_5$ space in the limit of 
$S\gg \sqrt{\lambda}$ scales as 
\be
\Delta-S={\sqrt{\lambda}\over \pi}\ln{S\over \sqrt{\lambda}}
+\cdots \;,
\ee
which looks the same as logarithmic growth of anomalous 
dimensions of operators with spin in the gauge theory. It has 
also been shown that the BMN  
operators \cite{BMN} can also be identified with classical
solutions of string in $AdS_5$ with angular momentum in $S^5$ 
space \cite{GKP}.

A generalization for the case when the string is stretched along
radial direction of $AdS$ and rotating 
along both $AdS_5$ and $S^5$ spaces has also been studied in 
\cite{FT}. This solution corresponds to those operators 
which have both spin and R-charge in the gauge theory side. 
A more general solution where the string is also
stretched in an angular coordinate of $S^5$ has been studied in 
\cite{Russo}. For further study in this direction see 
\cite{{BAR},{GSW}}.

The aim of this article is to generalize the classical 
string solution for M-theory
where we would have a membrane in the $AdS_7$ or $AdS_4$ 
background in M-theory. Regarding the fact that the AdS/CFT 
correspondence has also been conjectured for 
$AdS_7$ in M-theory which says that M-theory on 
$AdS_7\times S^4$ is dual to (0,2) theory, one would expect to 
find a classical rotating and boosted 
membrane solution in this background representing state with 
spin and R-charge in (0,2) theory. The same can be considered 
for $AdS_4\times S^7$

The organization of the paper is as following. In section 2, 
we shall review the results of \cite{{GKP},{FT}} considering the 
rotating and boosted string in the
$AdS_5\times S^5$ background. In this section we will rederive 
their solution using the Nambu action. In section 3, we will 
generalize this consideration for
$AdS_7\times S^4$ where we just have a Nambu-like action. 
The last section is devoted to discussion and some comments. 
We shall also make some comments for $AdS_4\times S^7$ case.  
  
\section{Review of closed string in $AdS_5\times S^5$}

In this section we shall review the results of \cite{GKP} 
considering a semi-classical solution of the rotating/boosted 
closed string stretched along the radial coordinate of 
$AdS_5\times S^5$ in type IIB 
string theory. A solution of rotating and 
boosted closed string has been studied in \cite{{FT},{Russo}}.

Let us start with the supergravity solution of $AdS_5\times S^5$ 
written in the global coordinates 
\bea
ds^2&=&R^2\bigg{[}-dt^2\cosh^2\rho+d\rho^2+\sinh^2\rho 
d\Omega_3^2 +d\psi_1^2+\cos^2\psi_1 (d\psi_2^2+
\cos^2\psi_2d{\tilde \Omega}_3^2)\bigg{]},\cr
&&\cr
d\Omega_3^2&=&d\beta_1^2+\cos^2\beta_1\;(d\beta_2^2+
\cos^2\beta_2d\beta_3^2),\cr
&&\cr
d{\tilde \Omega}_3^2&=&d\psi_3^2+\cos^2\psi_3\;
(d\psi_4^2+\cos^2\psi_4d\psi_5^2)
\label{MET5}
\eea
We would like to study a solution representing a rotating 
closed string configuration which is stretched along the 
radial coordinate. In order to study this system one needs 
to write an action for this closed string.
We note, however, that there are two different, but 
equivalent, ways to write the action for this string 
configuration; either we can use the Nambu
action or the Polyakov action. Since we are going to 
generalize this consideration to the M-theory membrane 
where we do not have the Polyakov-like action, we will
work with the Nambu action. This will give us an insight 
how to generalize the string calculations to the M-theory 
computations.

Now the point we should notice is the symmetries the 
string theory has. Let us parameterize the string worldsheet 
by $\sigma $ and $\tau$. Then we would have reparameterization 
invariance which has to be fixed. We can fix it by a 
parameterization such that the time coordinate of space-time, 
$t$ to be equal to worldsheet time, i.e. $t=\tau$. 
In this gauge a closed string configuration
representing a rotating string with angular velocity 
$\omega$ on $AdS_5$ space stretched along the radial 
coordinate is given by  
\be
t=\tau,\;\;\;\;\;\beta_3=\omega \tau,\;\;\;\;\;
\rho(\sigma)=\rho(\sigma+2\pi)
\label{CL}
\ee
all other coordinates are set to zero. For this solution 
the Nambu action,
\be
I=-{1\over 2\pi \alpha'}\int d\sigma^2\sqrt{-\det(G_{\mu\nu}
\partial_{a}X^{\mu}\partial_b X^{\nu})},
\label{NAMBUS}
\ee
reads
\be
I=-4{R^2\over 2\pi\alpha'}\int_0^{\rho_0}d\rho
\sqrt{{\dot t}^2\cosh^2\rho-
{\dot \beta}_3^2\sinh^2\rho}\;,
\ee
where dot represents derivative with respect to $\tau$. For our 
solution (\ref{CL}) ${\dot t}=1,\;{\dot \beta}_3=\omega$ and 
$\rho_0=\coth^{-1}(\omega)$. The factor of 4 comes from the 
fact that we are dealing with a folded closed string.
Working with one fold string, the string can be divided to 
four segments. Using the periodicity condition we just need 
to perform the integral for one quarter of string
multiplied by factor 4.

The two conserved momenta conjugate to $t$ and $\beta_3$ 
are the space-time energy $E$ and spin $S$. Using the above Nambu 
action these quantities are given by
\bea
E&=&4{R^2\over 2\pi\alpha'}\int_0^{\rho_0}
d\rho\frac{\cosh^2\rho}{\sqrt{\cosh^2\rho-\omega^2
\sinh^2\rho}}\;,\cr &&\cr S&=&4{R^2\omega\over
2\pi\alpha'}\int_0^{\rho_0}d\rho\frac{\sinh^2\rho}
{\sqrt{\cosh^2\rho-\omega^2\sinh^2\rho}}\;,
\label{ES5}
\eea
From the integrals (\ref{ES5}) one can proceed to compute 
the relation between energy and spin. To do this we can use 
an approximation in which the string
is much shorter or much longer than the radius of curvature 
of $AdS_5$. The short and long string limits are given by the 
limit in which $\rho_0\rightarrow 0$ and $\rho_0\rightarrow 
\infty$, respectively. 

\vspace*{.3cm}

{\bf Short strings}

\vspace*{.3cm}

For large $\omega$ one finds $\rho_0\sim {1\over \omega}
\rightarrow 0$ and therefore the string is much shorter than 
the radius of curvature of $AdS_5$. In fact in this
limit the $AdS_5$ space can be approximated by a flat metric 
near the center and therefore the calculation reduces to 
spinning string in the flat space. In this limit the integral
(\ref{ES5}) can be performed, and we find 
\be
E={R^2\over \omega\alpha'},\;\;\;\;\;\;\;
S={R^2\over 2\omega^2\alpha'}
\ee
so that
\be
E^2={2R^2\over\alpha'}\; S,
\ee
which is the well-known  flat space Regge trajectory.

\vspace*{.3cm}

{\bf Long strings}

\vspace*{.3cm}

On the other hand for $\omega\rightarrow 1$ the length 
of string scales as $\rho_0\sim {1\over 2}
\ln{2\over \omega-1}\rightarrow \infty$ and therefore
the string becomes long. In this case the spin is always 
large compare to the radius of curvature of $AdS_5$ space, 
i.e. $S\gg {R^2\over \alpha'}$. Setting 
$\rho_0={1\over 2}\ln{2\over \omega-1}$ in
(\ref{ES5}) one can find the approximate expansion of 
the integrals
\bea
E&=&{R^2\over 2\pi\alpha'}\left({2\over \omega-1}+
\ln{2\over \omega-1}+\cdots
\right)\;,\cr &&\cr
S&=&{R^2\over 2\pi\alpha'}\left({2\over \omega-1}-
\ln{2\over \omega-1}+\cdots
\right)\;,
\eea
which can be solved to find the dependence of energy 
on spin, which is
\be
E-S={R^2\over \pi\alpha'}\ln({\alpha'\over R^2}S).
\ee
Interesting enough this looks very similar to the logarithmic 
growth of anomalous dimensions of operators with spin in 
the gauge theory.

One can also consider a string rotating in $AdS_5 $ and 
in $S^5$ with independent angular velocity parameters 
$\omega$ and $\nu$ which is stretched along the radial 
coordinate $\rho$ from $\rho=0$ up to some 
$\rho=\rho_{max}\equiv\rho_0$. This closed string 
configuration is given by
\be
t=\kappa\tau,\;\;\;\;\beta_3=\omega\tau,\;\;\;\;
\psi_5=\nu\tau,\;\;\;\;
\rho(\sigma)=\rho(\sigma+2\pi).
\label{STCON}
\ee
This rotating and boosted closed string configuration 
has been studied in \cite{FT} where the authors used 
the Polyakov action. Now we would like to 
review their solution but with the Nambu action. 
This could give an insight how to generalize the 
situation for the M-theory case. Note that in the 
framework where the Nambu action is used we should  
impose the gauge fixing condition as following
\be
G_{\mu\nu}{\partial X^{\mu}\over \partial\tau}
{\partial X^{\nu}\over \partial\sigma}
=0,\;\;\;\;\;\;G_{\mu\nu}\left({\partial X^{\mu}\over 
\partial\tau}{\partial X^{\nu}\over \partial\tau}+
{\partial X^{\mu}\over \partial\sigma}{
\partial X^{\nu}\over \partial\sigma}\right)=0\;.
\label{GFIX}
\ee
We note that setting $t=\kappa\tau$ can not fix the diffeomorphism  
completely and in fact there is still a freedom to redefine the
parameters which are involved in the solution (\ref{STCON}). Imposing
the gauge fixing condition (\ref{GFIX}) would fix this freedom by
given a relation between them.
The first relation in equation (\ref{GFIX}) is automatically
satisfied for our string configuration (\ref{STCON}), 
while the second one leads to the following 
first order equation for $\rho$ 
\be
\left({d\rho\over d\sigma}\right)^2=
(\kappa^2-\nu^2)\cosh^2\rho-(\omega^2-\nu^2)\sinh^2\rho\;.
\label{RFIX}
\ee
The periodicity condition on $\rho$ is satisfied by 
considering a folded string. The string is folded onto 
itself, and the interval $0\leq \sigma < 2\pi$ is split into 
four segments. The function $\rho(\sigma)$ increases from 
zero to its maximal value which is given
 by setting ${d\rho\over d\sigma}({\pi\over 2})=0$, i.e.
\be
\cosh^2\rho_0={\omega^2-\nu^2\over \kappa^2-\nu^2}\;.
\ee
For the solution (\ref{STCON}) the Nambu action 
(\ref{NAMBUS}) reads
\be
I=-4{R^2\over 2\pi\alpha'}\int d\rho\sqrt{(\kappa^2-\nu^2)
\cosh^2\rho -(\omega^2-\nu^2)\sinh^2\rho}\;.
\ee
Using this action one can find the energy, spin and 
R-charge as following
\be
E=-{\partial I\over \partial \kappa},\;\;\;\;\;\;
S={\partial I\over \partial \omega},\;\;\;\;\;\;
J={\partial S\over \partial \nu},
\label{FORM}
\ee
which are 
\bea
E&=&{4R^2\over 2\pi\alpha'}\kappa\int_{0}^{\rho_0} 
\frac{\cosh^2\rho\;d\rho}{\sqrt{(\kappa^2-\nu^2)
\cosh^2\rho-(\omega^2-\nu^2)\sinh^2\rho}}\;,\cr &&\cr
S&=&{4R^2\over 2\pi\alpha'}\omega\int_{0}^{\rho_0} 
\frac{\sinh^2\rho\;d\rho}{\sqrt{(\kappa^2-\nu^2)
\cosh^2\rho-(\omega^2-\nu^2)\sinh^2\rho}}\;,\cr &&\cr
J&=&{4R^2\over 2\pi\alpha'}\nu\int_{0}^{\rho_0} 
\frac{d\rho}{\sqrt{(\kappa^2-\nu^2)\cosh^2\rho-
(\omega^2-\nu^2)\sinh^2\rho}}.
\label{ESJ5}
\eea
One observes that 
\be
E={\kappa\over \nu}J+{\kappa\over \omega}S\;,
\label{ESJ}
\ee
and the periodicity also implies an other condition on the
parameters
\be
2\pi=\int_0^{2\pi}d\sigma=4\int_0^{\rho_0}\frac{d\rho}{
\sqrt{(\kappa^2-\nu^2)\cosh^2\rho-(\omega^2-\nu^2)
\sinh^2\rho}}\;.
\ee
In particular one finds $J={R^2\over \alpha'}\nu$. This 
condition together with equations (\ref{ESJ}) and 
(\ref{ESJ5}) can be used to obtain the dependence of 
energy $E$ on spin and R-charge. The reader is 
refereed to \cite{FT} for detail of this computations.

\section{Rotating and boosted membrane in M-theory 
$AdS$ backgrounds}

Following \cite{GKP} we would like to consider the 
semi-classical solution of {\it closed membrane} in the 
M-theory representing states with higher
angular momentum on the $AdS$ and sphere parts which can 
be identified with {\it spin} and R-charge. 

Here we shall study the boosted and rotating 
{\it closed membrane} in the $AdS_7\times S^4$ generalizing 
the previously studied case for type IIB on 
$AdS_5\times S^5$ background \cite{{GKP},{FT},{Russo}}. 

Let us start with the gravity solution of 
$AdS_7\times S^4$ in the global coordinates 
\bea
l_p^{-2}dS^2&=& 4R^2\bigg{[}-dt^2\cosh^2\rho+d\rho^2+
\sinh^2\rho\left(d\psi_1^2+\cos^2\psi_1\; d\psi_2^2+
\sin^2\psi_1\;d\Omega_3^2\right)\cr && \cr
&+&{1\over 4}\left(d\alpha^2+\cos^2\alpha\;d\theta^2+
\sin^2\alpha\;(d\beta^2+\cos^2\beta\; d\gamma^2)\right)
\bigg{]}\;, \cr  &&  \cr
d\Omega^2_3&=&d\psi_3^2+\cos^2\psi_3\; d\psi_4^2+
\cos^2\psi_3\;\cos^2\psi_4\;d\psi_5^2,
\label{ADSBAC}
\eea
where $R^3=\pi N$. Note that we are using a unit in which 
$R$ is dimensionless.

The supersymmetric action of the M-theory
supermembrane has been studied in \cite{BST}. Here we 
shall only consider the bosonic part of the action 
which can be written as following
\be
I=-{1\over (2\pi)^2l_p^3}\int d\xi^3\left(
\sqrt{-\det(G_{\mu\nu}
\partial_{a}x^{\mu}\partial_b x^{\nu})}+
{1\over 6}\epsilon^{ijk}
\partial_ix^{\mu}\partial_jx^{\nu}\partial_k x^{\lambda}
C_{\mu\nu\lambda}
\right),
\label{M2ACT}
\ee
where $(\xi_1,\xi_2,\xi_3)=(\tau,\delta,\sigma)$ are 
coordinates which
parameterize the membrane worldvolume. $x^{\mu},\;\mu=0,
\cdots,10$ are
space-time coordinates and $C_{\mu\nu\lambda}$ is 
the massless M-theory three form. 

We look for a soliton solution corresponding to a closed 
membrane configuration in the background (\ref{ADSBAC}). 
We note that there are different membranes which could be 
considered depending on in which directions their worldvolume 
are taken. However it turns out that not all of these 
configurations are supersymmetric. Here we shall only 
consider the supersymmetric configuration of membrane. 
This can be given by a membrane rotating in $AdS_7$ and 
$S^4$ with independent angular velocity parameters 
$\omega$ and $\nu$ and is also stretched along the radial 
coordinate and another angular coordinate of $d\Omega_5$ 
part of $AdS_7$ space. This solution is given by\footnote{
The rotating membranes in the matrix model has been studied
in \cite{Bak}. For supermembrane solution on PP-wave background
ses also \cite{KY}.}

\be\ba {lll}
t=\kappa\tau\;, & \psi_2=\sqrt{2}a\delta\;, & \psi_5=
\sqrt{2}\omega \tau\;,\cr \rho=\rho(\sigma)\;,& 
\theta=2\nu\tau\;, & \psi_1={\pi\over 4}\;,
\label{MEMSOL}
\ea\ee
all other coordinates are set to zero. We have also 
periodicity condition for $\rho$ coordinate as 
$\rho(\sigma+2\pi)=\rho(\sigma)$. Therefore we are considering
a membrane configuration which is folded onto itself, and
the interval $0\leq \sigma < 2\pi$ is split into four 
segments. The function $\rho(\sigma)$ increases from zero 
to its maximal value, $\rho_0$  which is given by 
${d\rho\over d\sigma}=0$. We note also that, the same as 
$AdS_5$ in  the previous section, we should  
impose the gauge fixing condition which for bosonic 
membrane action are given by 
\be
G_{\mu\nu}{\partial X^{\mu}\over \partial \tau}
{\partial X^{\nu}\over \partial \sigma}=0,\;\;\;\;\;\;
G_{\mu\nu}{\partial X^{\mu}\over \partial \tau}
{\partial X^{\nu}\over \partial \delta}=0,
\label{C1}
\ee
and
\be
L^2G_{\mu\nu}{\partial X^{\mu}\over \partial \tau}
{\partial X^{\nu}\over \partial \tau}=\left(
G_{\mu\nu}{\partial X^{\mu}\over \partial \delta}
{\partial X^{\nu}\over \partial \sigma}\right)^2
-\left(G_{\mu\nu}{\partial X^{\mu}\over \partial \sigma}
{\partial X^{\nu}\over \partial \sigma}\right)
\left(G_{\lambda\gamma}
{\partial X^{\lambda}\over \partial \delta}
{\partial X^{\gamma}\over \partial \delta}\right),
\label{C2}
\ee
where $L$ is a constant with dimension of length. 
For our solution (\ref{MEMSOL}) $L=2R$ and moreover 
the constraints (\ref{C1}) are automatically satisfied 
while the last one, (\ref{C2}), leads to the following 
first order equation for $\rho$
\be
\left({d\rho\over d\sigma}\right)^2=
\frac{(\kappa^2-\nu^2)\cosh^2\rho-
(\omega^2-\nu^2)\sinh^2\rho}{a^2\sinh^2\rho},
\ee
which can be solved for $\sigma$. In fact from periodicity 
condition of $\rho$ we find
\be
2\pi=\int_0^{2\pi}d\sigma=4a\int_0^{\rho_0}\frac{\sinh\rho\; d\rho}
{\sqrt{(\kappa^2-\nu^2)\cosh^2\rho-
(\omega^2-\nu^2)\sinh^2\rho}}\;,
\ee
and moreover we rescaled $\delta$ such that ${4a\over 2\pi}=1$. 

For the solution (\ref{MEMSOL}) the CS part of the membrane 
action (\ref{M2ACT}) is zero and therefore the membrane action 
(\ref{M2ACT}) reads
\be
I=-{(2R)^3\over (2\pi)^2}a\int d\tau\; d\delta\; d\rho\; 
\sinh\rho \sqrt{\kappa^2\cosh^2\rho-\omega^2\sinh^2\rho-\nu^2}.
\ee 
From (\ref{FORM}) we can find the energy, spin and R-charge 
of the membrane as following
\bea
E&=&{4R^3\over \pi}\;\kappa\int_{0}^{\rho_0} 
\frac{\sinh\rho\;\cosh^2\rho\; d\rho}
{\sqrt{(\kappa^2-\nu^2)
\cosh^2\rho-(\omega^2-\nu^2)\sinh^2\rho}}\;,\cr &&\cr
S&=&{4R^3\over \pi}\;\omega\int_{0}^{\rho_0} 
\frac{\sinh^3\rho\;d\rho}{\sqrt{(\kappa^2-\nu^2)
\cosh^2\rho-(\omega^2-\nu^2)\sinh^2\rho}}\;,\cr &&\cr
J&=&{4R^3\over \pi}\;\nu\int_{0}^{\rho_0} 
\frac{\sinh\rho\;d\rho}{\sqrt{(\kappa^2-\nu^2)
\cosh^2\rho-(\omega^2-\nu^2)\sinh^2\rho}}\;,
\label{ESJ7}
\eea
where $\rho_0$ is 
\be
\rho_0=\coth^{-1}\sqrt{\frac{\omega^2-\nu^2}{\kappa^2-\nu^2}}.
\ee
The same as $AdS_5$ case we have
\be
E={\kappa\over \nu}J+{\kappa\over \omega}S\;,
\ee
which may be used to determine the dependence of $E$ on 
$S$ and $J$. The factor 4 has the same origin as one in 
$AdS_5$ in the previous section. An immediate consequence
of periodicity condition is that $J=\alpha_0\nu$ with 
$\alpha_0=4R^3/\pi$.

Now we have all ingredients to proceed to find the 
dependence of $E$ on the $S$ and $J$. To do this we 
need to perform the integrals (\ref{ESJ7}) which are 
involved in our study. In fact we get 
\bea
S&=&\alpha_0\frac{\omega}{\sqrt{\kappa^2-\nu^2}}
\;\frac{1}{2\eta^{3/2}}
\left[\sqrt{\eta}+(\eta-1)\left(\tan^{-1}\sqrt{\eta}-
{\pi\over 2}\right)\right]
\;,\cr &&\cr
J&=&\alpha_0\frac{\nu}{\sqrt{\kappa^2-\nu^2}}\;
{1\over \sqrt{\eta}}
\left({\pi\over 2}-\tan^{-1}
\sqrt{\eta}\right)\;,
\label{END}
\eea
where $\eta^{-1}=\sinh^2\rho_0$. These equations together 
with $J=\alpha_0\nu$ and $E=\kappa \alpha_0+
{\kappa\over \omega}S$ 
are enough to find the relation between 
$E, S$ and $J$. Following \cite{GKP} one can consider 
the limit of {\it short} 
$(\rho_0\rightarrow 0$ or $\eta\rightarrow \infty$), 
and {\it long} ($\rho_0\rightarrow \infty$ or 
$\eta\rightarrow 0$) membranes.

\vspace*{.4cm}

{\bf Short membranes}

\vspace*{.4cm}

For $\eta\rightarrow \infty$ we get
\be
\kappa^2\approx\nu^2+{1\over \eta^2},\;\;\;\;\;\;
\omega^2\approx \nu^2+{1\over \eta},
\ee
and from (\ref{END}) one finds
\be
S\approx {2\over 3}\alpha_0{1\over \eta}\;
\sqrt{\nu^2+{1\over \eta}}\;.
\ee
For the case of $\nu\ll {1\over \eta}$  we can solve the 
above equation for $\eta$
\be
{1\over \eta^{3/2}}\approx{3\over 2}\alpha_0^{-1} S\;,
\ee
and using the relation between $E$ and $S$ we get
\be
E=\alpha_0\sqrt{\nu^2+{1\over \eta^2}}+\cdots\;\;\;
\Longrightarrow\;\;\;\;\;
E\approx \alpha_0\;{1\over \eta}\;,
\ee
or
\be
E\approx (\frac{9}{\pi})^{1/3}R S^{2/3}.
\label{REGG}
\ee
Physically, in the short membrane limit, $\rho_0
\rightarrow 0$, the membrane is not stretched
much compared to the radius of curvature of $AdS_7$, 
therefore
we can approximate $AdS_7$ by flat metric near center.
In this case the calculation reduces to the rotating 
membrane in the flat space and in fact (\ref{REGG}) 
is analogues to the string in flat space where we get 
Regge trajectory, $E\sim \sqrt{S}$.

On the other hand for ${1\over \eta}\ll \nu \ll 
{1\over \sqrt{\eta}}$ one finds
\be
E=J+({\pi^2\over 24})^{1/3}R^{-2}\; JS^{2/3}+\cdots\;,
\ee
while for $\nu\gg {1\over \sqrt{\eta}}$ we get
\be
E=J+S+{3\over 2\pi^2}R^6\;{S^2\over J^3}+\cdots\;.
\label{PEN}
\ee

\vspace*{.4cm}

{\bf Long membranes}

\vspace*{.4cm}

In the long membranes limit where $\eta\rightarrow 0$ 
one finds
\be
\kappa^2\approx \nu^2+{\pi^2\over 4\eta},\;\;\;\;\;\;
\omega^2\approx \nu^2+{\pi^2\over 4}+{\pi^2\over 4\eta}\;,
\ee
moreover
\be
S\approx 
{\alpha_0\over 2}\;{1\over \eta}\sqrt{\nu^2+{\pi^2\over 4}+
{\pi^2\over 4\eta}}\;.
\label{SL}
\ee
Therefore the energy will be given by
\be
E\approx\alpha_0\sqrt{\nu^2+{\pi^2\over 4\eta}}+
\frac{\sqrt{\nu^2+{\pi^2\over 4\eta}}}
{\sqrt{\nu^2+{\pi^2\over 4}+{\pi^2\over 4\eta}}}\;S\;.
\ee
Note that in the long membranes limit the spin is always large 
compare to the radius of curvature of $AdS_7$ space. In this limit 
one can expand the above expression for energy. Using 
(\ref{SL}) we get the following dependence of energy on 
spin and R-charge
\bea
E-S&=&3S\left[{1\over 2}\left(\frac{R^3}{S}\right)^{2/3}+
{1\over 8} \left(\frac{R^3}{S}\right)^{4/3}+\cdots\right]
+{J\over 4}\bigg{[}{1\over 2}\frac{J}{R^3}
\left(\frac{R^3}{S}\right)^{1/3}\cr &&\cr
&+&\frac{J}{R^3}
\left(\frac{R^3}{S}\right)^{2/3}-
{1\over 16} \left(\frac{J}{R^3}
\right)^3\left(\frac{R^3}{S}\right)^{2/3}+
\cdots\bigg{]}\;.
\eea
On the other hand for the limit in which the angular 
momentum on $S^4$ is very small, $\nu\rightarrow 0$, 
the above expansion reads
\be
E-S=3S\left[{1\over 2}\left(\frac{R^3}{S}\right)^{2/3}+
{1\over 8} \left(\frac{R^3}{S}\right)^{4/3}+\cdots\right]\;,
\ee
which can be thought as a perturbative expansion with 
the expansion parameter $R\over S^{1/3}$. In fact in this 
limit a general form of the $S$ dependence of $E$ can be 
written as following
\be
E=S\left[2\left({R\over S^{1/3}}\right)^2+
{1\over \sqrt{1+\left({R\over S^{1/3}}\right)^2}}
\right]= S\sum_{n=0}^{\infty} c_n\left(\frac{R}
{S^{1/3}}\right)^{2n}\;,
\ee
where $c_n$ is some numerical factor. This is our 
prediction for the relation between energy and spin. 
Unfortunately our knowledge about (0,2) theory which is
conjectured to be dual to the M-theory in this background 
\cite{MAL} is too little to check this expression in 
the (0,2) theory side.

For the case of ${\pi\over 2}\ll \nu \ll {\pi\over 2\eta}$ 
we get
\be
E-S=2R^2 S^{1/3}+{1\over 4}
R^{-1}{J^2\over S^{2/3}}-{1\over 2^6}R^{-6}
{J^4\over S}\;,
\ee
while for the limit of $\nu\gg {\pi\over 2\eta}$ one finds
\be
E=J+S+2R^{4}{S^{2/3}\over J}-2R^{6}{S\over J^2}\;.
\ee

\section{Conclusions}

In this paper we have considered semi-classical 
membrane solution in the M-theory on $AdS_7\times S^4$ 
background representing a membrane rotating in $AdS_7$ and 
$S^4$ spaces with independent angular velocity 
parameters. We have studied the short and long membranes 
limits where we have obtained the dependence of energy,$E$, 
on spin, $S$, and R-charge, $J$. We have also shown that
for the case of $J\rightarrow 0$ 
while in the flat space limit the energy is proportional to 
$\sqrt{S^3}$ which is analogues to the Regge 
trajectory in string theory, in the long membrane limit 
where the effects of $AdS_7$ are important the $E-S$ at 
the leading order grows as $S^{1/3}$. In fact we have been 
able to write a closed form for the energy in terms of 
spin. This would be a prediction for the relation 
between energy and spin of 
the operators with large spin in (0,2) theory.   

An other interesting limit we have considered is the 
limit in which a short membrane is boosted
with the speed of light on $S^4$ space. The energy 
as a function of spin and R-charge is given by (\ref{PEN}). 
Indeed this corresponds to the case
where we are taking the Penrose limit of $AdS_7\times S^4$ 
\cite{BMN}. For comparison with the result of \cite{BMN} 
we could, for example, consider the one loop approximation 
around the classical solution (\ref{MEMSOL}) with $\omega=0$.
Although in this paper we have used the Nambu action all the
times, it seems using this action  for the one loop 
approximation is problematic. This is very similar to the
string case \cite{FT} where the induced metric had
singularity. In the string case the authors of
\cite{FT} could get rid of this problem using the
Polyakov action. But since in our case we have no such
an action one might wonder how to find a proper action 
for the fluctuations around our membrane solution. A way to 
do this could be to write the matrix model from 
membrane action. Then we can compare this matrix model with
one considered in \cite{BMN} (see also \cite{DSV}). To do this
we should first fix the light-cone gauge. In the light-cone gauge
the classical membrane solution (\ref{MEMSOL}) is given in terms of 
$x^+$ and $x^-$, 
which is $x^{+}=2\nu\tau,\;\;\;\;\; x^{-}=0$. 
Then we need to consider small fluctuations around the shrinking
closed membrane
\be
x^{+}=2\nu\tau+{{\tilde x}^+\over R},\;\;\;\;\rho={{\tilde\rho}
\over R},\;\;\;\;
x^{-}={{\tilde x^-}\over R^2},\;\;\;\;{\zeta}_i={{\tilde\zeta}_i
\over R}, 
\label{MMM}
\ee
where $\zeta_i$ stands for other coordinates . It can be shown that 
the light-cone
Hamiltonian (see \cite{deWitt}) in the quadratic approximation for 
bosonic membrane solution given by (\ref{MMM}) in the limit of 
$R\rightarrow \infty$ 
reduces to the light-cone Hamiltonian of a membrane in the PP-wave 
background \cite{DSV}.  

In this paper we have considered the case in which the 
membranes are only stretched along the radial coordinate. 
But the same as \cite{Russo} one could
consider a more general solution where the membranes are 
also stretched along an angular coordinate in $S^4$ part.

Similarly, one could also study the classical membrane solution 
in $AdS_4\times S^7$ background of M-theory. The gravity solution 
of $AdS_4\times S^7$ in the global 
coordinate is given by
\bea
ds^2={R^2\over 4}&\bigg{[}&-dt^2\cosh^2\rho+d\rho^2+
\sinh^2\rho\;(d\psi_1^2+
\cos^2\psi_1\;d\psi_2^2)\cr &&\cr 
&&+4(d\alpha^2+\cos^2\alpha\;d\Omega^2_3+\sin^2\alpha\;
d{\tilde \Omega}_3^2)\;\bigg{]},\cr &&\cr
d\Omega_3^2&=&d\beta_1^2+\cos^2\beta_1\;(d\beta_2^2+
\cos^2\beta_2\;d\beta_3^2)\cr&&\cr
d{\tilde \Omega}_3^2&=&d\beta_4^2+\cos^2\beta_4\;
(d\beta_5^2+\cos^2\beta_5\;d\beta_6^2)\;.
\eea
For the classical membrane solution representing a boosted membrane
on $S^7$ which is stretched along the radial coordinate and $\psi_1$
we will get the same results as $AdS_7$. This can be understood from 
the fact that the Penrose limits of $AdS_4\times S^7$ and 
$AdS_7\times S^4$ spaces give the same 11-dimensional PP-wave 
\cite{{Blau2},{BMN}}. 

{\bf Note added}: After submiting the paper, we were informed by 
P. Sundell that the rotating membrane in $AdS_7\times S^4$ has also
been studied in \cite{SS}. In fact their solution is an special case
of (\ref{MEMSOL}) in which $\kappa=1$ and $\nu=0$. We note, however,
that
for the case of $\nu=0$ setting $\kappa=1$ is enough to fix the gauge 
while for a general form one should further impose the gauge fixing
conditions (\ref{C1}) and (\ref{C2}). For the similar situation in 
string theory see \cite{GKP} and \cite{{FT},{Russo}}.

\vspace*{.5cm}

{ \bf Acknowledgements}

We would like to thank F. Ardalan, H. Arfaei, 
S. Parvizi, J. Russo, M. Sheikh-Jabbari, A. Tseytlin, H. 
Yavartanoo and L. Pando Zayas for useful comments 
and discussions.

\end{document}